\documentclass[english,aps,prl,showpacs,superscriptaddress,preprint]{revtex4-1}
\usepackage[english]{babel}
\usepackage[T1]{fontenc}
\usepackage{amsmath,amssymb,graphicx,mathrsfs,xcolor,babel,textcomp,esint}
\usepackage[unicode=true,pdfusetitle,bookmarks=true,bookmarksnumbered=false,bookmarksopen=false,breaklinks=true,pdfborder={0 0 0},backref=false,colorlinks=true]{hyperref}

\begin{document}

\title{Ultrafast Mapping of Coherent Dynamics and Density Matrix
Reconstruction in Terahertz-Assisted Laser Field}

\author{Yizhu Zhang}
\affiliation{Shanghai Advanced Research Institute, Chinese Academy of Sciences, Shanghai 201210, China}
\affiliation{Center for Terahertz waves and College of Precision Instrument and Optoelectronics Engineering, Key Laboratory of Opto-electronics Information and Technical Science, Ministry of Education, Tianjin University, China}
\author{Tian-Min Yan}
\email{yantm@sari.ac.cn}
\affiliation{Shanghai Advanced Research Institute, Chinese Academy of Sciences, Shanghai 201210, China}
\author{Y. H. Jiang}
\email{jiangyh@sari.ac.cn}
\affiliation{Shanghai Advanced Research Institute, Chinese Academy of Sciences, Shanghai 201210, China}
\affiliation{University of Chinese Academy of Sciences, Beijing 100049, China}
\affiliation{ShanghaiTech University, Shanghai 201210, China}

\begin{abstract}
  A time-resolved spectroscopic protocol exploiting terahertz-assisted
  photoionization is proposed to reconstruct transient density matrix.
  Population and coherence elements are effectively mapped onto spectrally
  separated peaks in photoionization spectra. The beatings of coherence
  dynamics can be temporally resolved beyond the pulse duration, and the
  relative phase between involved states is directly readable from the
  oscillatory spectral distribution. As demonstrated by a photo-excited
  multilevel open quantum system, the method shows potential applications for
  sub-femtosecond time-resolved measurements of coherent dynamics with free
  electron lasers and tabletop laser fields.
\end{abstract}
\maketitle

Quantum coherence derived from the principle of superposition is a fundamental
concept in quantum mechanics and a ubiquitous phenomenon with the time scale
ranging from milliseconds, as in delicately prepared cold atoms, to
attoseconds for electronic coherence in field-perturbed atoms
{\cite{pabst_decoherence_2011}} and molecules {\cite{arnold_electronic_2017}}.
Although the coherence dynamics take a key role in photo-reactions
{\cite{engel_evidence_2007,collini_coherent_2009,collini_coherently_2010}},
the real-time observation of electronic coherence is highly challenging,
particularly on the femtosecond or even attosecond timescale. In
photo-ionization processes in atoms and molecules, the electronic coherence
prepared by shaking the inner-shell electrons has been uncovered to lose
within $\sim$1 fs. The decoherence is closely related to the correlation
between the photoelectron and the parent ion, significantly affecting the
formation of coherent hole wave packets
{\cite{pabst_decoherence_2011,arnold_electronic_2017}}. In addition, for the
recently investigated photosynthetic complexes, the possible mechanism of
coherence-assisted energy transfer is explored by the multi-dimensional
optical spectroscopy {\cite{engel_evidence_2007}}. Due to the involvement of
multiple transport pathways in the extremely complicated biomolecular
environment, further supporting evidences from experimental observations are
still required. The observation of ultrafast electronic coherence dynamics is
indispensable to provide valuable insight into these fundamental processes.
Despite of the significance, ultrafast coherence dynamics are not readily
obtained from commonly used time-resolved (pump-probe) spectroscopies, since
the comparatively large population signal usually conceals the signature of
the coherence.

The complete dynamic information of an open quantum system is encoded in the
time-evolved density matrix elements $\rho_{i  j} (\tau)$, including the
population $\rho_{i  i}$ and the coherence $\rho_{i  j}$ $(i \neq j)$. The
$\rho_{i  j} (\tau)$-reconstruction from dedicated designed experiments, e.g.,
the quantum process tomography with the multi-dimensional spectroscopy to
fully describe a quantum black box {\cite{yuen-zhou_quantum_2011}}, is always
desirable. The temporal resolution of optical spectroscopy, however, is
temporally restricted to tens of femtoseconds. Since the quantum states are
typically prepared and probed by femtosecond-duration pulses, the convolution
with the probing pulse leads to the smear of signatures of faster electronic
coherences.

In this sense, observing the ultrafast electronic coherence in atoms and
molecules, in principle, requires the attosecond metrology. Exploiting the
attosecond transient absorption spectroscopy, the electronic coherence of the
valence electron in krypton ions was measured with the sub-femtosecond
temporal accuracy {\cite{goulielmakis_real-time_2010}}. The same spectroscopic
methodology also allows observing the correlated two-electron coherence motion
in helium with attosecond temporal resolution
{\cite{ott_reconstruction_2014}}. Although the attosecond metrology can probe
the electron wave-packet motion with high temporal resolution, it requires
tremendous efforts with sophisticated laser techniques, including the precise
control of both the amplitude and phase of the femtosecond light field
throughout the measurement. Moreover, contributions from population and
coherence are overlapped after one-dimensional spectral projections, thus
usually concealing quantum coherence. For the recently developed
free-electron-laser (FEL) facilities, the compression of the pulse with high
energetic photons (from extreme ultraviolet to hard X-ray region) down to the
sub-femtosecond scale is difficult, obstructing further insights into
ultrafast electronic coherence. Kowalewski {{\em et al\/}}. proposed to record
the sub-femtosecond electronic coherence using a femtosecond probe pulse
{\cite{kowalewski_monitoring_2016}}. The photoelectron is driven by a
phase-locked near infrared (NIR) pulse and multiple sidebands around the
characteristic peaks are created. The sidebands carry the high-resolution
information of coherence, yet simultaneously complicate the spectroscopic
analysis due to the possibly contaminated characteristic spectral features.
Moreover, the prerequisite that the photon energy of the streaking field has
to be resonant with the relevant states may become a restriction.

In this work, we propose a spectroscopic method to retrieve the full
information of the density matrix, like the two-dimensional spectroscopy,
unraveling the contributions of coherence in an extra dimension. Exploiting
the femtosecond extreme ultraviolet (XUV) pulse and terahertz (THz) streaking
field, we show density matrix elements directly map into the photoelectron
spectra. Especially, the method in principle allows observing the motion of
the electronic wave packet in FEL facilities. In contrast to proposals
operating in the so called "sideband regime"
{\cite{kowalewski_monitoring_2016}}, our method works in the ``streaking
regime'' {\cite{helml_measuring_2014}}. Moreover, without the restriction that
the photon energy must match the energy gap between coherently excited states
{\cite{kowalewski_monitoring_2016}}, the method can be applied in systems
without {{\em a priori\/}} knowledge. Conventionally, the streaking technique
is used to characterize the temporal profile of ultrashort pulses, e.g., the
attosecond streaking technique that calibrates an attosecond pulse using a
moderately strong NIR light field {\cite{kienberger_steering_2002}}, the FEL
calibration using the THz or IR-streaking pulse
{\cite{fruhling_single-shot_2009,grguras_ultrafast_2012,helml_measuring_2014}}.
Here, instead of field diagnosis, the THz-streaking technique is used to
investigate ultrafast dynamics.

\begin{figure}[h]
  \includegraphics[scale=1.5]{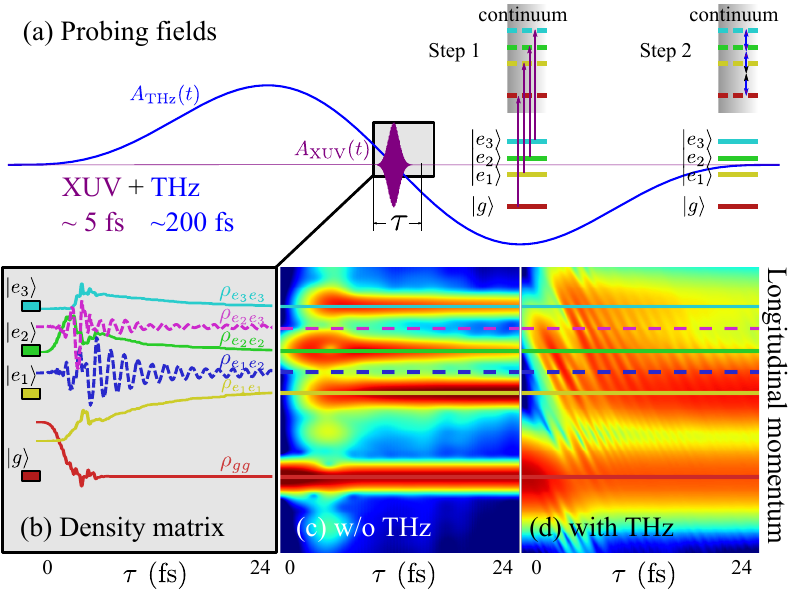}
  \caption{\label{fig:scheme}Schematic of the THz streaking-assisted
  photo-ionization experiment that allows for the real-time density matrix
  imaging. (a) The probing pulse configuration in the measurement. To observe
  the unknown quantum dynamics of an open quantum system within the temporal
  window (gray box), a femtosecond XUV pulse (purple) with a well-synchronized
  THz streaking pulse (blue) as a probe beam scans over time delay $\tau$. The
  XUV pulse is locked at the zero-crossing of the THz vector potential. (b)
  The energy levels and the time evolution of the density matrix elements
  $\rho_{i  j}$ within the time window. (c) Photoelectron momentum
  distribution without THz field. Populations $\rho_{i  i}$ are mapped into
  spectral peaks of different longitudinal momentum $p$, as indicated by
  colors of solid lines. The $\tau$-dependent spectral yields, labeled by red,
  yellow, green and cyan lines for $\rho_{g g}$, $\rho_{e_1 e_1}$, $\rho_{e_2
  e_2}$ and $\rho_{e_3 e_3}$, respectively, reveal the evolution of
  populations. (d) Photoelectron momentum distribution with THz streaking.
  Spectral peaks are broadened and the emergent oscillating fringes, labeled
  along blue and purple dashed lines, can be shown to account for coherences
  $\rho_{e_1 e_2}$ and $\rho_{e_2 e_3}$, respectively. The fringe can be well
  resolved, though the duration of the probing XUV pulse is much longer than
  its period of oscillation. The effective separation of $\rho_{i  j}$ along
  $p$ and the temporally resolved coherence dynamics allow one to reconstruct
  the time evolution of $\rho_{i  j}$.}
\end{figure}

For a multilevel open quantum system, the scheme of the streaking-assisted
photo-ionization is illustrated in Fig. \ref{fig:scheme}(a). The probe fields,
as proposed in this work, comprise a femtosecond XUV pulse and a well
synchronized THz pulse, the zero-crossing of whose vector potential is
required to be locked at the center of the XUV pulse. As illustrated by the
inset of Fig. \ref{fig:scheme}(a), the combination of the two pulses realizes
a two-step process. First, the single-photon ionization induced by the XUV
pulse liberates the electron of the superposed state at $\tau$, taking a
frozen snapshot of quantum states exactly at that time. The THz wave then
drives the photoelectron, kinetically rearranging electronic trajectories and
inducing the interference between trajectories from different states in the
final momentum distribution $w (p ; \tau)$, which is a critical step to the
observation of the coherence. We will show that, the dynamics of the system at
any time $\tau$ as described by $\rho_{i  j} (\tau)$ [Fig.
\ref{fig:scheme}(b)] are mapped directly onto $w (p ; \tau)$, with different
elements differentiated by longitudinal momentum $p$. Scanning over $\tau$,
the measured $w (p ; \tau)$, as shown in Fig. \ref{fig:scheme}(c) and (d),
allows for reconstructing the time evolved $\rho_{i  j} (\tau)$. Without the
THz field [Fig. \ref{fig:scheme}(c)], spectral peaks depict the time-evolved
populations, which can also be measured using other time-resolved techniques.
With the THz field [Fig. \ref{fig:scheme}(d)], interference fringes are formed
upon broadened peaks, which is of essential concern due to the encoded
information of coherence. All details of these figures can be referred to Fig.
\ref{fig:open-system}. The further analysis shows that $w (p ; \tau)$ is
composed of a series of Gaussian peaks corresponding to all $\rho_{i  j}
(\tau)$, laying the foundation for $\rho_{i  j} (\tau)$-reconstruction. The
method provides an alternative tool to investigate the decoherence of
attosecond photo-ionization in atoms and molecules, and the coherence energy
transport in complex photo-reaction systems.

The $w (p ; \tau)$ to measure in the experiment can be analyzed under the
strong field approximation. Assuming the atom is subject to linearly polarized
fields, $E (t) = E_{\mathrm{XUV}} (t) + E_{\mathrm{THz}} (t)$, the ionization
signal $w (p ; \tau) = |p| |M_p (\tau) |^2$ is collected along the
polarization direction. Neglecting the high-order above threshold ionization,
the amplitude of the direct ionization in the length gauge reads $M_p (\tau) =
\lim_{t \rightarrow \infty} \int_0^t \mathrm{d} t' \mathrm{e}^{\mathrm{i} S_p
(t')} \langle p + A (t') | \hat{\mu} E (t') | \Psi_0 (t') \rangle$, where $A
(t) = A_{\text{XUV}} (t) + A_{\text{THz}} (t)$ is the vector potential and
$S_p (t') = \frac{1}{2}  \int_0^{t'} \mathrm{d} t''  [p + A (t'')]^2$. Note,
$\tau$ is hidden as the temporal center of pulses in $A (t)$ and $E (t)$.
$\hat{\mu}$ is the transition dipole, and $| \Psi_0 (t) \rangle$ is the
initial (superposition) state. The photoelectron emission process consists of
two steps. The laser field firstly liberates the electron from the atom when
the XUV pulse plays a dominant role. Subsequently, the photoelectron is
drifted to the final momentum. Since the amplitudes $\left| A_{\mathrm{XUV}}
\right| \ll \left| A_{\mathrm{THz}} \right|$, the free electron is driven
dominantly by the THz electric field. According to the analysis of the
ionization probability in the streaking regime (see S.I), when amplitudes of
the wave functions vary slowly within the temporal window of the XUV pulse,
the distribution $| M_p (\tau) |^2$ comprises Gaussian peaks for all
population and coherence terms, $| M_p (\tau) |^2 = \frac{\pi
(E_0^{(\ensuremath{\operatorname{XUV}})})^2}{2 | b (p) |} \left[
W_{\text{pop}} (p ; \tau) + W_{\text{coh}} (p ; \tau) \right]$, with
\begin{eqnarray}
  W_{\text{pop}} (p ; \tau) & = &  \sum_i \rho_{i  i} (\tau) \mathrm{e}^{-
  \frac{\Omega_{i  i}^2 (p)}{| b (p) \sigma |^2}},  \label{eq:w-population}\\
  W_{\text{coh}} (p ; \tau) & = & 2 \sum_{i, j < i} \text{Re} \left[ \rho_{i 
  j} (\tau) \mathrm{e}^{\mathrm{i} \frac{\alpha p \Omega_{i  j} (p) \Delta_{i 
  j}}{| b (p) |^2}} \right] \nonumber\\
  &  & \times \mathrm{e}^{- \frac{\Omega_{i  j}^2 (p)}{| b (p) \sigma |^2}}
  \mathrm{e}^{- \frac{(\Delta_{i  j} / 2)^2}{| b (p) \sigma |^2}}, 
  \label{eq:w-coherence}
\end{eqnarray}
from which the positions and widths of spectral peaks are easily read. Here,
$\Omega_{i  j} (p) = p^2 / 2 + \left( I_{\text{p}}^{(i)} + I_{\text{p}}^{(j)}
\right) / 2 - \omega_{\text{XUV}}$, where $\omega_{\mathrm{XUV}}$ is the
photon energy of XUV field and $I_{\text{p}}^{(i)} = - E_i$ is the ionization
potential of the $i$th state. Spectral peaks appear at $p$ where $\Omega_{i 
j} (p) = 0$, whose solutions include $p_{i  i} = \pm \sqrt{2 \left(
\omega_{\text{XUV}} - I_{\text{p}}^{(i)} \right)}$ for $\rho_{i  i}$, and
$p_{i  j} = \pm \sqrt{2 \left( \omega_{\text{XUV}} - \left( I_{\text{p}}^{(i)}
+ I_{\text{p}}^{(j)} \right) / 2 \right)}$ that is exactly between $p_{i  i}$
and $p_{j  j}$ for $\rho_{i  j}$ $(i \neq j)$. Thus, $\rho_{i  j} (\tau)$ in
principle can be inferred from $w (p_{i  j} ; \tau)$ at characteristic
momentum $p_{i  j}$. The widths of peaks for both $W_{\text{pop}}$ and
$W_{\text{coh}}$ are determined by $| b (p) \sigma | = \sqrt{1 / \sigma^2 +
(\alpha p \sigma)^2}$ with $b (p) = 1 / \sigma^2 - \mathrm{i} \alpha p$
dependent on the THz field amplitude $\alpha$ and the temporal width of XUV
field $\sigma$ (standard deviation). In Eq. (\ref{eq:w-coherence}), though
$W_{\text{coh}}$ is much smaller than $W_{\text{pop}}$ due to the presence of
the last exponential term of $\Delta_{i  j} = I_{\text{p}}^{(i)} -
I_{\text{p}}^{(j)}$, the relative amplitude of $W_{\text{coh}}$ enhances when
the THz field increases. For the energetic spaces spanning over a few eV in
typical molecular transitions, the THz electric field of $\sim$kV/cm, which is
conveniently provided by tabletop THz sources nowadays, is sufficient to
couple the coherently excited states and reveal $W_{\text{coh}}$.

In the neighborhood of $p_{i  j}$, the exponent in the real part of Eq.
(\ref{eq:w-coherence}), $p \Omega_{i  j} (p) / | b (p) |^2 \simeq p_{i  j}^2 
(p - p_{i  j}) / (1 / \sigma^4 + \alpha^2 p_{i  j}^2)$, is linearly
proportional to $p$. Defining $\varphi_{i  j} (\tau) = \Delta_{i  j} \tau +
\varphi_i - \varphi_j$ the instantaneous relative phase between states $i$ and
$j$, whose initial phases are $\varphi_i$ and $\varphi_j$, respectively, the
real part in Eq. (\ref{eq:w-coherence}) becomes $| \rho_{i  j} (\tau) | \cos
[- \Delta_{i  j} \alpha p_{i  j}^2  (p - p_{i  j}) / (1 / \sigma^4 + \alpha^2
p_{i  j}^2) + \varphi_{i  j} (\tau)]$, indicating that the spectral yield
oscillates with both $\tau$ and $p$. As expected, the oscillatory frequency
$\Delta_{i  j}$ along $\tau$-axis agrees with the evolution of $\rho_{i  j}
(\tau)$ $(i \neq j)$, which can be resolved beyond the duration of the XUV
pulse. The oscillation along $p$-axis, however, is noteworthy, since it
provides a potential single-$\tau$ approach to reconstruct phase $\varphi_{i 
j} (\tau)$ without the necessity of scanning over time delay $\tau$.

Taking the simplest example of a two-level system, we show the feasibility of
coherence imaging and relative phase reconstruction using the THz-streaking
method. Assuming the system is prepared in the superposition state with the
equal probabilities, $| \Psi_0 (t) \rangle = (|g \rangle \mathrm{e}^{-
\mathrm{i} E_g  (t - t_0) + \mathrm{i} \varphi_g} + |e \rangle \mathrm{e}^{-
\mathrm{i} E_e  (t - t_0) + \mathrm{i} \varphi_e}) / \sqrt{2}$, with $|g
\rangle$ and $|e \rangle$ the ground state and excited state of energies $E_g$
and $E_e$, respectively. Here we consider a system with $E_g = - 13.6$ eV and
$E_e = - 3.4$ eV, and initial phases are $\varphi_e = \varphi_g = 0$ at the
initial time $t_0 = 0$ fs. The photon energy of the XUV pulse is 2 a.u. (54.4
eV) with the peak field amplitude 0.005 a.u. (intensity $\sim 9 \times
10^{11}$ W/cm$^2$), and the duration is 5 fs. The center frequency of the
single-cycle THz pulse is 4 THz and the peak field strength is 0.001 a.u.
($\sim 5$ MV/cm). In principle, the requirement that XUV pulse is temporally
locked at the zero-crossing of $A_{\mathrm{THz}} (t)$ can be satisfied in the
FEL facility, where the XUV and THz pulses, generated by the same electron
bunch from the linear accelerator, are well synchronized
{\cite{fruhling_single-shot_2009}}. The jitter between the excitation pulse
and probing FEL pulse is hardly controlled within attosecond accuracy
nowadays, but the restriction can be overcome by the continuously developed
technique of synchronization. Since there is no need for the single-$\tau$
measurement, the reliable statistics of spectra can be provided with
state-of-art synchronization of the pulse sequence. In our approach, the
Fourier transform limited XUV pulse is required. Although Fourier
transformation limited pulse is not conveniently delivered in the
self-amplified spontaneous emission (SASE) mode, the seeding operation
improves the temporal coherence and the profile of FEL pulse
{\cite{ackermann_generation_2013}}.

\begin{figure}[h]
  \includegraphics[scale=1.5]{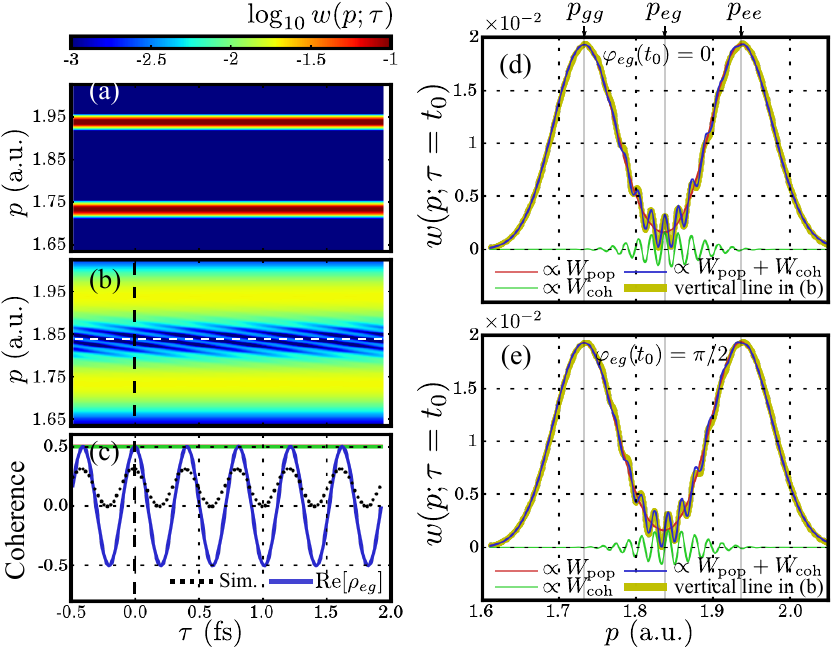}
  \caption{\label{fig:closed-system}The temporally resolved quantum coherence
  with streaking-assisted photo-ionization. In a two-level system, we assume
  the amplitudes of the two states are constant with initial relative phase
  $\varphi_{e  g} (t_0) = \varphi_{e } - \varphi_g = 0$ at $t_0 = 0$ fs.
  Panels (a) and (b) show $w (p ; \tau)$ without and with THz-streaking,
  respectively. Panel (c) shows both the coherence
  $\mathrm{\ensuremath{\operatorname{Re}}} [\rho_{e  g} (\tau)]$ (blue) and
  the temporal profile of the simulated spectral signal $w (p = p_{e  g} ;
  \tau)$ (black dotted line), as extracted along the white dashed line in (b).
  Panel (d) presents $w (p ; \tau = t_0)$ as extracted along the black dashed
  line in (b). The extracted result (yellow) and the sum over model-based
  Gaussian peaks (blue), including the contributions from $W_{\text{pop}}$ in
  Eq. (\ref{eq:w-population}) (red) and $W_{\text{coh}}$ in Eq.
  (\ref{eq:w-coherence}) (green), are compared. The relative phase $\varphi_{e
  g} (t_0)$ can be reconstructed with a single-$\tau$ measurement at time
  $\tau = t_0$ simply by examining $w (p ; t_0)$ along the $p$-axis. When
  $\varphi_{e  g} (t_0) = 0$, the coherence (green) at $p = p_{e  g}$ is
  maximized as $\cos (\varphi_{e  g} (t_0)) = 1$. By comparison, assuming
  $\varphi_{e  g} (t_0) = \pi / 2$, the distribution is shown in (e), where
  the coherence at $p = p_{e  g}$ becomes 0.}
\end{figure}

\begin{figure}[h]
  \includegraphics[scale=1.5]{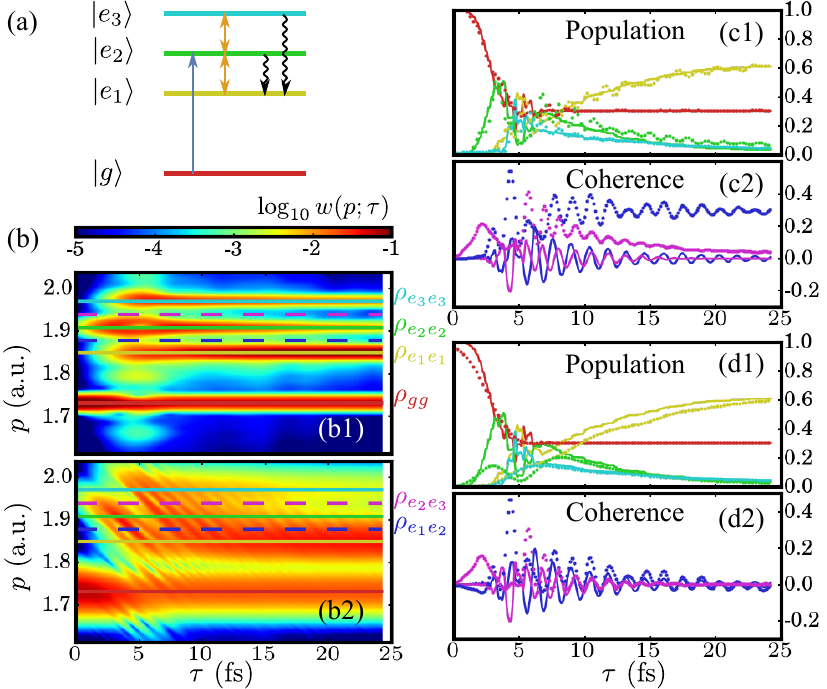}
  \caption{\label{fig:open-system}Reconstruction of $\rho_{i  j} (\tau)$ in a
  multilevel open system. The four-level system under the excitation of
  two-color fields is shown in (a) with energies $(E_g, E_{e_1}, E_{e_2},
  E_{e_3}) = (- 0.5, - 0.29, - 0.18, - 0.06)$ a.u.. With the two-color pulse
  sequence, coherences between these states are created. The relaxation
  processes are assumed to occur within the excited manifold, where the
  population in $|e_2 \rangle$ and $|e_3 \rangle$ may jump randomly to the
  lowest excited state $|e_1 \rangle$ with constant rates. Panels (b1) and
  (b2) show the photoelectron momentum distributions without and with
  streaking-THz field, respectively. The peaks of characteristic momenta $p_{i
  j}$, along which the results are extracted for $\rho_{i 
  j}$-reconstruction, are indicated by dashed lines of the same color code as
  used for energy levels in (a). In (c), the THz-streaking signals extracted
  from (b2) (dotted lines with the same color code) are compared with the true
  $\rho_{i  j}$ (solid lines). Panel (d) shows the $\rho_{i  j} (\tau)$
  reconstructed with the protocol.}
\end{figure}

Scanning the probing fields over $\tau$ yields $w (p ; \tau)$ as shown in Fig.
\ref{fig:closed-system}. Panels (a) and (b) show the spectrograms before and
after applying the streaking THz field, respectively. As the kinetic energy of
the photoelectron from the $i$th state is $\omega_{\mathrm{XUV}} -
I_{\text{p}}^{(i)}$, the spectral peaks at $p_{g  g} = 1.73$ a.u. and $p_{e 
e} = 1.94$ a.u. originate from states $|g \rangle$ and $|e \rangle$,
respectively. The two $\tau$-independent peaks recover $W_{\text{pop}}$ in Eq.
(\ref{eq:w-population}) since the amplitudes of $|g \rangle$ and $|e \rangle$
are constants. With the THz field, the two peaks in panel (b) are
significantly broadened, and in the intermediate region around $p_{e  g} =
1.84$ a.u. fringes are formed. Extracting $w (p = p_{e  g} ; \tau)$, as
indicated by the white dashed line in Fig. \ref{fig:closed-system}(b), the
fringes shown in Fig. \ref{fig:closed-system}(c) oscillate with the period 0.4
fs, corresponding to 10.2 eV between states $|g \rangle$ and $|e \rangle$, in
good agreement to the oscillatory pattern of coherence
$\mathrm{\ensuremath{\operatorname{Re}}} [\rho_{eg} (\tau)]$. Remarkably, the
THz-streaking assisted method allows for resolving the sub-femtosecond quantum
beating that cannot be achieved by a 5-fs probing XUV pulse alone. On the
other hand, as depicted by Eq. (\ref{eq:w-coherence}), Fig.
\ref{fig:closed-system}(d) for $w (p ; \tau = t_0)$ also shows the oscillatory
behavior with $p$ near $p = p_{e  g}$, in good agreement with the sum over
analytically derived Gaussian peaks. The $p$-dependent oscillation allows for
the easy reconstruction of relative phase at arbitrary time $\tau$. Comparing
with $\varphi_{e  g} (t_0) = 0$ in (d), the distribution for $\varphi_{e  g}
(t_0) = \pi / 2$ presented in (e) indicates that $\varphi_{e  g} (t_0)$ can be
directly mapped by the phase of oscillatory $w (p ; \tau = t_0)$ at $p = p_{e 
g}$. Note, the fringes near $p_{e  g}$ in (b) are not merely caused by the
superposition of broadened peak of $W_{\text{pop}}$ at $p_{e   e}$ and $p_{g 
g}$. According to Eq. (\ref{eq:w-coherence}), there indeed exists the
inbetween peak at $p_{e  g}$ for coherence $W_{\text{coh}} (p_{e  g})$ even
without THz field, but its relative intensity can be significantly enhanced by
the streaking-THz field.

The simple two-level system, as shown in Fig. \ref{fig:closed-system}, has
demonstrated that $\rho_{i  j} (\tau)$ are well embedded in the
streaking-assisted $w (p ; \tau)$. However, the method is particularly useful
when complicated energy structure is involved and the ultrafast relaxation and
decoherence dynamics in open quantum systems are of concern. Here, we propose
a reconstruction protocol (see S.III) to extract $\rho_{i  j} (\tau)$ in an
ensemble consisting of four-level atoms that interact with near-resonant
pulses [Fig. \ref{fig:open-system}(a)]. Since the photoionization is evaluated
based on the wave function $| \Psi_0 (t') \rangle$, the open quantum system is
simulated using the method of Monte Carlo wave function (quantum jump, see
S.IV) {\cite{dalibard_wave-function_1992}}, which is equivalent to the
Lindblad-type master equation, allowing for the description of spontaneous
emission and decoherence.

The THz-free and THz-streaking measurements are shown in Fig.
\ref{fig:open-system}(b1) and (b2), respectively. Due to the densely spaced
energy levels, the relatively weak THz field at 0.0006 a.u. (2 MV/cm) is
sufficient to signify $w (p_{e_1 e_2} ; \tau)$ and $w (p_{e_2 e_3} ; \tau)$
for coherences between adjacent excited states. All other parameters of the
probing fields are the same as used for the two-level system. Note that in
Fig. \ref{fig:open-system}(b2) with spectral peaks dramatically broadened by
the THz field, beating patterns along $\tau$ are all over the spectrogram,
including $w (p_{i  i} ; \tau)$ $(i = g, e_1, e_2, e_3)$ within the excited
manifold. Extracting the simulated result for populations along dashed lines
in Fig. \ref{fig:open-system}(b2) at $p = p_{i  i}$, these artificial
oscillations are clearly seen in Fig. \ref{fig:open-system}(c1). In addition,
in Fig. \ref{fig:open-system}(c2) for coherences, the extracted results
significantly deviate from the true $\rho_{i  j} (\tau)$ $(i \neq j)$. All
these deviations suggest additional post-process be required to obtain the
correct $\rho_{i  j} (\tau)$. Therefore, we introduce the reconstruction
protocol utilizing both the THz-free and THz-streaking $w (p ; \tau)$ to
effectively eliminate the contributions from surrounding peaks (see S.III).
Then, after applying quantum jump correction (only slightly influences results
in our situation, the detailes will be discussed in the subsequent work), the
reconstructed $\rho_{i  j} (\tau)$ are shown in Fig. \ref{fig:open-system}(d)
comparing with the true $\rho_{i  j} (\tau)$. Except for the poor agreement
around $t_0$ as resolving the fast excitation dynamics is restricted by the
relatively large XUV pulse duration, the reconstructed $\rho_{i  j} (\tau)$
from the simulated measurement well recovers the true $\rho_{i  j} (\tau)$,
validating the feasibility of density matrix reconstruction using
streaking-assisted photoelectron spectra.

In summary, a spectroscopic protocol is proposed to fully describe the
evolution of density matrix elements, which can be reconstructed from
streaking-assisted photoelectron spectrum. With the XUV pulse of fs-scale
width, the beating feature of the quantum coherence can be retrieved with the
sub-fs precision, less restricted by the duration of probing field than in
conventional transient spectroscopies. The analysis of the streaking process
is substantially simplified by a model with the spectral distribution
described by the sum over Gaussian peaks, whose characteristic momenta
effectively separate all density matrix elements. The model also indicates the
photoelectronic spectrum oscillates with $p$ near the characteristic momentum
$p_{i  j}$ for coherence, providing an innovative single-$\tau$ approach to
phase reconstruction between quantum states at arbitrary time. Above all, the
systematic protocol of $\rho_{i  j} (\tau)$-reconstruction is demonstrated by
a simulated measurement observing the excitation and decoherence dynamics in a
multilevel open system. The method offers the possibility to observe the
sub-femtosecond decoherence of attosecond photo-ionization in small atomic and
molecular systems. For more complex systems, e.g., biological molecules and
complexes, the coherent energy transfer in photosynthetic complexes may also
be investigated.

The study was supported by National Natural Science Foundation of China (NSFC)
(11420101003, 61675213, 11604347, 91636105), Shanghai Sailing Program
(16YF1412600).

\bibliographystyle{apsrev4-1}
\bibliography{main}

\appendix

\section{Analysis of the streaking-assisted ionization}

It is assumed that the atom is subject to linearly polarized fields and the
momentum distribution of photoelectrons, $w (p) = | p | | M_p |^2$, is only
collected along the polarization direction. The amplitude of the direct
ionization with the strong field approxiamtion (SFA) in the length gauge
reads,
\begin{eqnarray}
  M_p & = & \lim_{t \rightarrow \infty} \int^t \mathrm{d} t'
  \mathrm{e}^{\mathrm{i} S_p (t')} \langle p + A (t') | \hat{\mu} E (t') |
  \Psi_0 (t') \rangle  \label{eq:Mp-sfa}
\end{eqnarray}
where $S_p (t') = \frac{1}{2} \int^{t'} \mathrm{d} t''  [p + A (t'')]^2$. The
initial state, $| \Psi_0 (t') \rangle$, may be prepared as the superposition
of states $\{ | \psi_i \rangle \}$ of the ionization energy
$I_{\text{p}}^{(i)}$, $| \Psi_0 (t') \rangle = \sum_i c_i (t')
\mathrm{e}^{\mathrm{i} I_{\text{p}}^{(i)} t'} | \psi_i \rangle$. Substituting
the superposition into Eq. (\ref{eq:Mp-sfa}),
\begin{eqnarray}
  M_p & = & \lim_{t \rightarrow \infty} \sum_i \int^t \mathrm{d} t' c_i (t')
  \mu_i [k (t')] E (t') \mathrm{e}^{\mathrm{i} S_{p, I \text{p}}^{(i)} (t')}, 
  \label{eq:Mp-sum}
\end{eqnarray}
with $k (t') = p + A (t')$, the dipole matrix element $\mu_i (k) \equiv
\langle k| \hat{\mu} | \psi_i \rangle$ and the action $S_{p,
I_{\text{p}}}^{(i)} (t') = \int^{t'} \mathrm{d} t''  \left[ k (t'')^2 / 2 +
I_{\text{p}}^{(i)} \right]$. Defining $M_p^{(i)}$ the integral in Eq.
(\ref{eq:Mp-sum}), we recast $M_p = \sum_i M_p^{(i)}$ a superposition of
transition amplitudes from all components of initial states.

\subsubsection{Partial contribution from $M_p^{(i)}$}

In the following we focus on the partial contribution $M_p^{(i)}$ from the
channel of initial state $i$. When only the XUV pulse is present, $E (t') =
E_{\ensuremath{\operatorname{XUV}}} (t')$ and $A (t') =
A_{\ensuremath{\operatorname{XUV}}} (t')$. Since the amplitude
$A_0^{(\ensuremath{\operatorname{XUV}})} =
E^{(\ensuremath{\operatorname{XUV}})}_0 /
\omega_{\ensuremath{\operatorname{XUV}}}$ is relatively small as
$\omega_{\ensuremath{\operatorname{XUV}}}$ is large, the action reduces to
$S_{p, I_{\text{p}}}^{(i)} (t') \simeq \int^{t'} \left[ \frac{1}{2} p^2 +
I_{\text{p}}^{(i)} \right] \mathrm{d} t'' = \left( \frac{1}{2} p^2 +
I_{\text{p}}^{(i)} \right) t'$. Defining $E_{\ensuremath{\operatorname{XUV}}}
(t') =\mathcal{E}_{\ensuremath{\operatorname{XUV}}} (t' - \tau) \cos
[\omega_{\ensuremath{\operatorname{XUV}}}  (t' - \tau)]$ with
$\mathcal{E}_{\ensuremath{\operatorname{XUV}}} (t' - \tau)$ the pulse envelope
centered at $\tau$, the phase in the exponential reads $\frac{p^2}{2} +
I_{\text{p}}^{(i)} \pm \omega_{\ensuremath{\operatorname{XUV}}}$, and the fast
oscillating part (the "$+$" term) can be neglected after the integration.
Also neglecting the dipole term, the transition amplitude is approximately
\begin{eqnarray}
  M_p^{(i)} (\tau) & \sim & \lim_{t \rightarrow \infty} \frac{1}{2}
  \mathrm{e}^{\mathrm{i} \left( \frac{p^2}{2} + I_{\text{p}}^{(i)} \right)
  \tau} \int^t_{- \infty} \mathrm{d} t' c_i (t')
  \mathcal{E}_{\ensuremath{\operatorname{XUV}}} (t' - \tau)
  \mathrm{e}^{\mathrm{i} \Omega_{i  i} (p)  (t' - \tau)}, 
  \label{eq:Mp-without-THz}
\end{eqnarray}
where $\Omega_{i  i} (p) = p^2 / 2 + I_{\text{p}}^{(i)} -
\omega_{\ensuremath{\operatorname{XUV}}}$. It can be viewed as the short time
Fourier transform of $c_i (t)$ with the window function
$\mathcal{E}_{\ensuremath{\operatorname{XUV}}} (t)$. The spectrum $| M_p^{(i)}
(\tau) |^2$ peaks at $p^2 / 2 = \omega_{\ensuremath{\operatorname{XUV}}} -
I_{\text{p}}^{(i)}$.

With the presence of both XUV and THz pulses, the conditions
$A_0^{(\ensuremath{\operatorname{XUV}})} \ll
A_0^{(\ensuremath{\operatorname{THz}})}$ and
$E_0^{(\ensuremath{\operatorname{XUV}})} >
E_0^{(\ensuremath{\operatorname{THz}})}$ justify the approximation $k (t')
\simeq p + A_{\text{THz}\text{}} (t')$. With the center time of the THz field
$\tau$ coincident with the XUV field, $E_{\text{THz}} (t')
=\mathcal{E}_{\text{$\ensuremath{\operatorname{THz}}$}} (t' - \tau) \cos
\left[ \omega_{\text{THz}} (t' - \tau) \right]$, since the duration of the XUV
pulse is much smaller than the THz field, the latter is approximately linear,
i.e., $A_{\ensuremath{\operatorname{THz}}} (t') \simeq \alpha (t' - \tau)$,
around the streaking time when ionization events occur. Conducting the series
expansion of $A_{\text{THz}} (t')$ around $\tau$, it is shown that the slope
$\alpha = - E_0^{(\ensuremath{\operatorname{THz}})}$. Substituting the linear
approximation of $A_{\ensuremath{\operatorname{THz}}} (t')$ into action $S_{p,
I_{\text{p}}}^{(i)} (t')$, we find $S_{p, I_{\text{p}}}^{(i)} (t') = \left[
\left( \frac{p^2}{2} + I_{\text{p}}^{(i)} \right) + \alpha p \left(
\frac{t'}{2} - \tau \right) + \frac{\alpha^2}{6}  (t^{\prime 2} - 3 t' \tau +
3 \tau^2) \right] t'$. The quadratic term of $\alpha$ is negligible when
$\alpha$ is small, thus $S_{p, I_{\text{p}}}^{(i)} (t') \simeq \frac{\alpha
p}{2} t^{\prime 2} + \left( \frac{p^2}{2} + I_{\text{p}}^{(i)} - \alpha p \tau
\right) t' = \frac{\alpha p}{2} (t' - \tau)^2 + \left( \frac{p^2}{2} +
I_{\text{p}}^{(i)} \right)  (t' - \tau) + \phi_i (p, \tau)$ with $\phi_i (p,
\tau) = \left( \frac{p^2}{2} + I_{\text{p}}^{(i)} \right) \tau - \left(
\frac{\alpha p}{2} \right) \tau^2$ a phase that can be factored out of the
integration over $t$. Similar to the analysis without the THz streaking,
substituting the XUV field, $E_{\ensuremath{\operatorname{XUV}}} (t')
=\mathcal{E}_{\ensuremath{\operatorname{XUV}}} (t' - \tau) \cos
[\omega_{\ensuremath{\operatorname{XUV}}}  (t' - \tau)]$, into $M_p^{(i)}
(\tau)$, the transition amplitude in the polarization direction is given by
\begin{eqnarray}
  M_p^{(i)} (\tau) & \sim & \frac{1}{2} e^{\mathrm{i} \phi_i (p, \tau)}
  \lim_{t \rightarrow \infty} \int^t_{- \infty} \mathrm{d} t' c_i (t')
  \mathcal{E}_{\ensuremath{\operatorname{XUV}}} (t' - \tau)
  \mathrm{e}^{\mathrm{i} \left[ \frac{\alpha p}{2}  (t' - \tau)^2 + \Omega_i
  (p)  (t' - \tau) \right]},  \label{eq:Mp-with-THz}
\end{eqnarray}
as long as the XUV field is within the linear region of the THz field.
Comparing with Eq. (\ref{eq:Mp-without-THz}), the appearance of an extra
quadratic term in the exponent, $(\alpha p / 2)  (t' - \tau)^2$, results in
the spectral broadening.

\subsubsection{Gaussian envelope of $\mathcal{E}_{\text{XUV}}$ and slowly
varying envelope approximation of $c_i (t')$}

Assuming the envelope is Gaussian,
$\mathcal{E}_{\ensuremath{\operatorname{XUV}}} (t' - \tau) =
E_0^{(\ensuremath{\operatorname{XUV}})} \mathrm{e}^{- (t' - \tau)^2 / 2
\sigma^2}$, we cast the integral part of Eq. (\ref{eq:Mp-with-THz}) as $I =
\int_{- \infty}^{\infty} \mathrm{d} t' c_i (t') \mathrm{e}^{- X (t' - Y_i)^2 +
C_i}$, where $X = b (p) / 2$, $Y_i = \tau + \mathrm{i} \Omega_{i  i} (p) / b
(p)$, $C_i = [\mathrm{i} \Omega_{i  i} (p)]^2 / 2 b (p)$ and $b (p) = 1 /
\sigma^2 - \mathrm{i} \alpha p$. Integrating by parts, it is shown that
\begin{eqnarray}
  I & = & \frac{\sqrt{\pi} \mathrm{e}^{C_i}}{2 \sqrt{X}} E_0^{\left(
  \text{XUV} \right)}  \left[ c_i (\tau + \Delta \tau) + c_i (\tau - \Delta
  \tau) - \int_{\tau - \Delta \tau}^{\tau + \Delta \tau} \mathrm{d} t' 
  \text{Erf} \left[ \sqrt{X} (t' - Y_i) \right]  \dot{c}_i (t') \right], 
  \label{eq:integral-I}
\end{eqnarray}
where $\text{Erf} [\cdot]$ is the error function, and $\Delta \tau$ is an
appropriate value to estimate the active temporal region of the XUV pulse. As
a rule of thumb, $\Delta \tau = 3 \sigma$ is sufficient to cover the main part
of the $\mathcal{E}_{\text{XUV}} (t' - \tau)$.

In Eq. (\ref{eq:integral-I}), the known properties of $c_i (t')$ may yield
further simplification. If $c_i (t')$ changes slowly in regards to $2 \Delta
\tau$, $\dot{c}_i (t') \Delta \tau / c_i (t') \ll 1$, Eq.
(\ref{eq:integral-I}) becomes $I = \frac{\sqrt{\pi}
\mathrm{e}^{C_i}}{\sqrt{X}}  \bar{c}_i (\tau)$ with $\bar{c}_i (\tau)$ the
averaged wave amplitude near $\tau$. In other words, if the XUV pulse is
sufficiently narrow, the wave amplitude $c_i (\tau)$ can always be well
resolved within the window. Then the ionization amplitude reads
\begin{eqnarray}
  M_p^{(i)} (\tau) & = & \frac{E_0^{\left( \text{XUV} \right)}}{2}  \bar{c}_i
  (\tau) \mathrm{e}^{\mathrm{i} \phi_i (p, \tau)} \sqrt{\frac{\pi}{X}}
  \mathrm{e}^{C_i} \nonumber\\
  & = & \frac{E_0^{(\ensuremath{\operatorname{XUV}})}}{2}  \bar{c}_i (\tau)
  \mathrm{e}^{\mathrm{i} \phi_i (p, \tau)}  \sqrt{\frac{2 \pi}{b (p)}}
  \mathrm{e}^{- \frac{\Omega_i^2 (p)}{2 b (p)}}  \label{eq:Mp-slow-c}
\end{eqnarray}

\subsubsection{Ionization probability and momentum distribution}

In the following, assuming the wave function varies slowly in regards to the
window width of the XUV pulse, we calculate the ionization signal by
substituting Eq. (\ref{eq:Mp-slow-c}), the partial contribution from channel
$i$, into $| M_p (\tau) |^2 = \left| \sum_i M_p^{(i)} (\tau) \right|^2$,
\begin{eqnarray*}
  | M_p (\tau) |^2 & = & \frac{\pi
  (E_0^{(\ensuremath{\operatorname{XUV}})})^2}{2 | b (p) |}  \left\{ \sum_i |
  c_i |^2 \mathrm{e}^{- \frac{\Omega_{i  i}^2 (p)}{| b (p) \sigma |^2}} +
  \sum_{i, j \neq i} (c_i^{\ast} c_j) \mathrm{e}^{- \mathrm{i} [\phi_i (p,
  \tau) - \phi_j (p, \tau)]} \mathrm{e}^{- \frac{1}{2}  \left[ \frac{\Omega_{i
  i}^2 (p)}{b^{\ast} (p)} + \frac{\Omega_{j  j}^2 (p)}{b (p)} \right]}
  \right\} .
\end{eqnarray*}
Here, $\bar{c}_i (\tau)$ is abbrivated by $c_i$. \ The exponent in the last
term that is associated to the contribution from the coherence can be further
reduced. First,
\begin{eqnarray*}
  \frac{\Omega_{i  i}^2 (p)}{b^{\ast} (p)} + \frac{\Omega_{j  j}^2 (p)}{b (p)}
  & = & \frac{\text{Re}  [b (p)]  [\Omega_{i  i}^2 (p) + \Omega_{j  j}^2 (p)]
  + \mathrm{i} \text{Im} [b (p)]  [\Omega_{i  i}^2 (p) - \Omega_{j  j}^2
  (p)]}{| b (p) |^2} .
\end{eqnarray*}
By substituting $b (p)$ and $\Omega_{i  i} (p)$, one can show in the above
equation, $\Omega_{i  i}^2 (p) + \Omega_{j  j}^2 (p) = 2 \Omega_{i  j}^2 (p) +
\Delta_{i  j}^2 / 2$ and $\Omega_{i  i}^2 (p) - \Omega_{j  j}^2 (p) = 2
\Omega_{i  j}^2 (p) \Delta_{i  j}$, where $\Delta_{i  j} = I_{\text{p}}^{(i)}
- I_{\text{p}}^{(j)}$ and $\Omega_{i  j} (p) = [\Omega_i (p) + \Omega_j (p)] /
2 = p^2 / 2 + \left( I_{\text{p}}^{(i)} + I_{\text{p}}^{(j)} \right) / 2 -
\omega_{\ensuremath{\operatorname{XUV}}}$ consistent with the definition of
$\Omega_{i  i} (p)$. Using $\ensuremath{\operatorname{Re}} [b (p)] = 1 /
\sigma^2$ and $\mathrm{\ensuremath{\operatorname{Im}}} [b (p)] = - \alpha p$,
we have
\begin{eqnarray*}
  \frac{\Omega_{i  i}^2 (p)}{b^{\ast} (p)} + \frac{\Omega_{j  j}^2 (p)}{b (p)}
  & = & \frac{2 \Omega_{i  j}^2 (p)}{| b (p) \sigma |^2} + \frac{\Delta^2_{i 
  j}}{2 | b (p) \sigma |^2} - \mathrm{i} \frac{2 \alpha p \Omega_{i  j} (p)
  \Delta_{i  j}}{| b (p) |^2} .
\end{eqnarray*}
Also note that in $| M_p |^2$ the phase difference $\phi_i (p, \tau) - \phi_j
(p, \tau) = \left( I_{\text{p}}^{(i)} - I_{\text{p}}^{(j)} \right) \tau =
\Delta_{i  j} \tau$. Therefore, the analytical expression of spectral peaks,
Eq. (1) and (2) in the main text, is derived, $| M_p (\tau) |^2 = \frac{\pi
(E_0^{(\ensuremath{\operatorname{XUV}})})^2}{2 | b (p) |}  \left\{
W_{\text{pop}} (p ; \tau) + W_{\text{coh}} (p ; \tau) \right\}$, where
\begin{eqnarray}
  W_{\text{pop}} (p ; \tau) & = & \sum_i \rho_{i  i} (\tau) \mathrm{e}^{-
  \frac{\Omega_{i  i}^2 (p)}{| b (p) \sigma |^2}},  \label{eq:W-pop}\\
  W_{\text{coh}} (p ; \tau) & = & \sum_{i, j \neq i} \rho_{i  j} (\tau)
  \mathrm{e}^{- \frac{\Omega_{i  j}^2 (p)}{| b (p) \sigma |^2}} \mathrm{e}^{-
  \frac{(\Delta_{i  j} / 2)^2}{| b (p) \sigma |^2}} \mathrm{e}^{\mathrm{i}
  \frac{\alpha p \Omega_{i  j} (p) \Delta_{i  j}}{| b (p) |^2}} \nonumber\\
  & = & 2 \sum_{i, j < i} \mathrm{\ensuremath{\operatorname{Re}}} \left[
  \rho_{i  j} (\tau) \mathrm{e}^{\mathrm{i} \frac{\alpha p \Omega_{i  j} (p)
  \Delta_{i  j}}{| b (p) |^2}} \right] \mathrm{e}^{- \frac{\Omega_{i  j}^2
  (p)}{| b (p) \sigma |^2}} \mathrm{e}^{- \frac{(\Delta_{i  j} / 2)^2}{| b (p)
  \sigma |^2}},  \label{eq:W-coh}
\end{eqnarray}
where $\rho_{i   j} (\tau) = (c_i^{\ast} c_j) \mathrm{e}^{- \mathrm{i}
\Delta_{i  j} \tau}$ are density matrix elements. Eq. (\ref{eq:W-pop}) and
(\ref{eq:W-coh}) present several features of the photoelectronic spectrum in
the streaking regime,
\begin{itemize}
  \item The exponential $\mathrm{e}^{- \Omega_{i  j}^2 (p) / | b (p) \sigma
  |^2}$ in Eq. (\ref{eq:W-pop}) and (\ref{eq:W-coh}) indicates that both
  $W_{\text{pop}} (p)$ and $W_{\text{coh}} (p)$ share the same spectral
  profile of Gaussian envelope. The spectral peak is located at $p = p_{i  j}$
  that satisfies $\Omega_{i  j} (p_{i  j}) = 0$, effectively mapping $\rho_{i 
  j} \rightarrow p_{i  j}$ and separating different density matrix elements in
  the spectrum. Note, the momentum $p_{i  j} (i \neq j)$ for coherence is
  between $p_{i  i}$ and $p_{j  j}$ for corresponding populations. The width
  of the spectral peak is determined by $| b (p) \sigma |^2 = 1 / \sigma^2 +
  (\alpha p \sigma)^2$. Without the THz pulse, the XUV pulse alone leads to
  the width of the spectral peak $\sim 1 / \sigma$, which is exactly the
  reciprocity relation between temporal and spectral domains and complies with
  the uncertainty principle. With the THz pulse, the spectral width can be
  effectively broadened by $\alpha$, the amplitude of streaking-THz field.
  
  \item $W_{\text{coh}} (p ; \tau)$ is typically much smaller than
  $W_{\text{pop}} (p ; \tau)$ due to factor $\mathrm{e}^{- (\Delta_{i  j} /
  2)^2 / | b (p) \sigma |^2}$. When $\Delta_{i  j}$ is large, it is unlikely
  to observe the peak of $W_{\text{coh}} (p ; \tau)$ clearly. However, the
  extra THz field helps alleviate the reduction caused by the factor and
  enhance $W_{\text{coh}} (p ; \tau)$. Therefore, the THz-streaking technique
  is particular useful when $\Delta_{i  j}$ is large and the coherence is of
  interest. On the other hand, if $\Delta_{i  j}$ is small, the coherence can
  have already been clearly seen even without THz field. It means, the
  strength of THz field should be optimized depending on the \ \
  
  \item Although the method can be used to study the multilevel system, the
  resolutions in temporal and spectral domains still follow uncertainty
  principle dependent on the XUV pulse width $\sigma$, i.e., one cannot
  achieve high spectral resolution and temporal resolution simultaneously. As
  shown by Eq. (\ref{eq:Mp-with-THz}), the time correlation between electric
  field $\mathcal{E} (\tau)$ of the finite pulse width and $c_i (\tau)$ smears
  the true $c_i (\tau)$. However, the beating patterns of
  $\mathrm{\ensuremath{\operatorname{Re}}} [\rho_{i  j} (\tau)]$ can be well
  temporally resolved beyond $\sigma$, as long as $W_{\text{coh}}$ is
  measurable.
  
  \item The factor $\mathrm{e}^{\mathrm{i} \frac{\alpha p \Omega_{i  j} (p)
  \Delta_{i  j}}{| b (p) |^2}}$ in (\ref{eq:W-coh}) also suggests the momentum
  distribution oscillates with $p$ around $p_{i  j}$. Increasing the THz
  strength may help observe the oscillation. Besides the phase reconstruction
  as mentioned in the main text, the frequency of the oscillation may also be
  used to calibrate the light field due to the dependence on $\alpha$ and
  $\omega_{\text{XUV}}$.
\end{itemize}

\section{Reconstruction of relative phase}

The relative phase can be extracted from photoelectronic spectrum. Assuming
that the initial phase is given by $\varphi_i$ for the $i$th state, $| \Psi_0
(t) \rangle = \sum_i c_i (t) \mathrm{e}^{\mathrm{i} \varphi_i}
\mathrm{e}^{\mathrm{i} I_{\text{p}}^{(i)} t} |i \rangle$ and $c_i (t) \in
\mathbb{R}$ is the amplitude of the wave function, the element $\rho_{i  j}$
can be represented by the relative phase embedded expression, $\rho_{i  j} (t)
= (c_i (t) \mathrm{e}^{\mathrm{i} \varphi_i})^{\ast}  (c_j (t)
\mathrm{e}^{\mathrm{i} \varphi_j}) \mathrm{e}^{- \mathrm{i} \Delta_{i  j} t} =
| \rho_{i  j} (t) | \mathrm{e}^{- \mathrm{i} (\varphi_i - \varphi_j)}
\mathrm{e}^{- \mathrm{i} \Delta_{i  j} t}$ with $| \rho_{i  j} (t) | = c_i (t)
c_j (t)$. Thus, the coherence part Eq. (\ref{eq:W-coh}) reads,
\begin{eqnarray*}
  W_{\text{coh}} (p ; \tau) & = & 2 \sum_{i, j < i} | \rho_{i  j} (\tau) | 
  \mathrm{\ensuremath{\operatorname{Re}}} \left[ (\mathrm{e}^{- \mathrm{i}
  (\varphi_i - \varphi_j)}  \mathrm{e}^{- \mathrm{i} \Delta_{i  j} \tau}
  \mathrm{e}^{\mathrm{i} \frac{\alpha p \Omega_{i  j} (p) \Delta_{i  j}}{| b
  (p) |^2}} \right] \mathrm{e}^{- \frac{\Omega_{i  j}^2 (p)}{| b (p) \sigma
  |^2}} \mathrm{e}^{- \frac{(\Delta_{i  j} / 2)^2}{| b (p) \sigma |^2}}\\
  & = & 2 \sum_{i, j < i} | \rho_{i  j} (\tau) | \cos \left[ \Delta_{i  j} 
  \left( \tau - \frac{\alpha p \Omega_{i  j} (p) }{| b (p) |^2} \right) +
  \varphi_i - \varphi_j \right] \mathrm{e}^{- \frac{\Omega_{i  j}^2 (p)}{| b
  (p) \sigma |^2}} \mathrm{e}^{- \frac{(\Delta_{i  j} / 2)^2}{| b (p) \sigma
  |^2}}
\end{eqnarray*}
Since near $p \simeq p_{i  j}$, the term $\frac{\alpha p \Omega_{i  j} (p)}{|
b (p) |^2} = \frac{\alpha p \left( \frac{p^2}{2} + \left( I_{\text{p}}^{(i)} +
I_{\text{p}}^{(j)} \right) / 2 - \omega_{\text{XUV}}
\right)}{\frac{1}{\sigma^4} + \alpha^2 p^2} \simeq \frac{\alpha p_{i  j}^2
}{\frac{1}{\sigma^4} + \alpha^2 p_{i  j}^2}  (p - p_{i  j})$ is approximately
linear to $p$, $W_{\text{coh}} (p ; \tau)$ reads
\begin{eqnarray*}
  W_{\text{coh}} (p ; \tau) & \simeq & 2 \sum_{i, j < i} | \rho_{i  j} (\tau)
  | \cos \left[ \Delta_{i  j}  \left( \tau - \frac{\alpha p_{i  j}^2
  }{\frac{1}{\sigma^4} + \alpha^2 p_{i  j}^2}  (p - p_{i  j}) \right) +
  \varphi_i - \varphi_j \right] \mathrm{e}^{- \frac{\Omega_{i  j}^2 (p)}{| b
  (p) \sigma |^2}} \mathrm{e}^{- \frac{(\Delta_{i  j} / 2)^2}{| b (p) \sigma
  |^2}}\\
  & = & 2 \sum_{i, j < i} | \rho_{i  j} (\tau) | \cos \left[ -
  \frac{\Delta_{i  j} \alpha p_{i  j}^2 }{\frac{1}{\sigma^4} + \alpha^2 p_{i 
  j}^2}  (p - p_{i  j}) + \varphi_{i  j} (\tau) \right] \mathrm{e}^{-
  \frac{\Omega_{i  j}^2 (p)}{| b (p) \sigma |^2}} \mathrm{e}^{-
  \frac{(\Delta_{i  j} / 2)^2}{| b (p) \sigma |^2}},
\end{eqnarray*}
where $\varphi_{i  j} (\tau) = \Delta_{i  j} \tau + (\varphi_i - \varphi_j)$
is the relative phase difference between states $i$ and $j$ at time $\tau$.
Hence, the relative phase difference for arbitrary time $\tau$ can be easily
inferred from the phase of $W_{\text{coh}} (p)$ at $p = p_{i  j}$, when the
cosine part becomes $\cos (\varphi_{i  j} (\tau))$.

Assuming that at a given time $\tau$, the initial relative phase between
state $i$ and $j$ is desired to be measured. According to the scheme in this
work, one only needs to move the center of the combinational fields to time
$\tau$, and retrieve the photoelectronic spectrum. It is expected the
THz-streaking field allows one to clearly observe the oscillation around $p =
p_{i  j}$, then determine the phase at exactly this point, from which the
relative quantum phase between two states can be evaluated. The advantage of
the method is that it only requires a single-$\tau$ measurement for the
desired time $\tau$ without further scanning procedure over $\tau$.

\section{Protocol of $\rho_{i  j}$-reconstruction }

With the information of the density matrix elements embedded in the
photoelectronic momentum distribution, a systematic protocol is desired to
quantitatively reconstruct the $\rho_{i  j}$. Since the spectrum in the
streaking-regime is described by model as the sum over Gaussian peaks, the
components can be disentangled for the $\rho_{i  j}$-reconstruction. The most
straightforward method is to consider the twice measurements. First, with only
XUV field, one is able to extract populations $\rho_{i  i}$. Then, applying
the THz field, the acquired coherence-encoded spectrum can be further
processed by the known $\rho_{i  i}$ to find out the coherence $\rho_{i  j}$.
The more detailed procedure is as following,

1. With the spectrum under only the XUV field, in order to derive $\rho_{i 
i}$, read $w (p_{i  i})$ versus $\tau$, from which factor out the coefficient
according to Eq. (\ref{eq:W-pop}).

2. With the spectrum under both the XUV and the THz fields, for the coherence
$\rho_{i  j}$, since the signal $w (p_{i  j})$ is alway positive [see, e.g.,
black line around $p_{e  g}$ in Fig. 2(d) in the main text] due to the
contributions from adjacent Gaussian peaks $W_{\text{pop}} (p_{i  j})$ [Eq.
(\ref{eq:W-pop})], one should substract these background components first,
then reconstruct $\rho_{i  j}$ using Eq. (\ref{eq:W-coh}).

However, the above scheme is applicable only when energy levels are well
distinguishable, i.e., $\rho_{i  i} (\tau)$ can be clearly identified. When
the energy levels are densely distributed, the THz-induced broadened spectral
peaks may interfer, even between Gaussian peaks of $W_{\text{pop}} (p)$,
leading to beating pattern of $w (p_{i  i})$ as the function of $\tau$,
eventually yielding artificial oscillatory behavior of $\rho_{i  i} (\tau)$
after the above steps. Therefore, it is practical to first conduct the
measurement without THz field, just as the conventional pump-probe experiment,
to retrieve $\rho_{i  i} (\tau)$ correctly. Of course, it is required at least
the XUV pulse alone is capable to resolve the energy levels. After that,
introducing the THz field of appropriate intensity that enhances the coherence
relevant signals $w (p_{i  j})$, one can reconstruct $\rho_{i  j} (\tau)$
using the previously obtained $\rho_{i  i} (\tau)$.

\section{Method of Monte Carlo wave function (MCWF)}

In Eq. (\ref{eq:Mp-sum}), the calculation of photoelectronic momentum
distribution requires that the coefficients $c_i (t)$ of all wave function are
known. For an open system, however, the ensemble consisting of many samples
that are different from each other has to be considered, e.g., spontaneous
emission occurs randomly among atoms in an ensemble. Such problems usually
requires solving the the master equation of density matrix, and the system
cannot be simply described by the wave function of pure state. However, with
$\rho_s$ the density matrix of the subsystem we are interested in and $C_m
(C_m^{\dagger})$ the lowering (raising) operator in the subsystem, it has been
shown the master equation with Lindblad-type relaxation operator $\mathcal{D}=
- \frac{1}{2} \sum_m (C_m^{\dagger} C_m \rho_s + \rho_s C_m^{\dagger} C_m) +
\sum_m C_m \rho_s C_m^{\dagger}$ can be equivalently solved by the MCWF. The
MCWF allows for the access to wave functions that are later used for the
calculation of the photoelectronic spectrum. Further details of the MCWF may
be referred to [J. Dalibard, Y. Castin, and K. M{\o}lmer, Phys. Rev. Lett. 68,
580 (1992)].

The example in the main text shows that the four level system initially in
the ground state $|g \rangle$ is excited by the external fields, while later
the transferred components in the highly excited state $|e_2 \rangle$ and
$|e_3 \rangle$ may spontaneously jump back to the lowest excited state $|e_1
\rangle$. Assuming the wave function of the system reads $| \Psi \rangle =
\sum_i c_i \mathrm{e}^{\mathrm{i} I_{\text{p}}^{(i)} t} |i \rangle$ with the
index of basis $i \in \{ g, e_1, e_2, e_3 \}$, the hamiltonian reads $\hat{H}
= \sum_i E_i |i \rangle \langle i| + \sum_{i, j} W (t) |i \rangle \langle j|$.
The off-diagonal terms, $W (t) = 0.02 \mathrm{e}^{- (t - 4100)^2 / 60^2} \sin
(0.324 t) + 0.04 \mathrm{e}^{- (t - 4200)^2 / 60^2} \sin (0.112 t)$, describe
interactions between the system with the two pulses of $\omega_1 = 0.324$
centered at 4100 a.u., and $\omega_2 = 0.112$ centered at 4200 a.u.. Note,
because of the long pulse duration of the THz field ($\sin^2$-envolope is used
in the model), the time center at 4000 a.u. is defined here as the time zero
$t_0$ for actual numerical calculation. Later for clear demonstration, $t_0 =
4000$ a.u. will be shifted to $t_0 = 0$ a.u. as presented in the main text.
The energies of four levels are $(E_g, E_{e_1}, E_{e_2}, E_{e_3}) = (- 0.5, -
0.29, - 0.18, - 0.06)$. Pulse 1 [blue line in Fig. 3(a)] is near resonant with
states $|g \rangle$ and $|e_2 \rangle$, effectively transfering population
from $|g \rangle$ to $|e_2 \rangle$. Pulse 2 [yellow line in Fig. 3(a)]
couples states within the excited manifold ($|e_1 \rangle$-$|e_2 \rangle$ and
$|e_2 \rangle$-$|e_3 \rangle$), further distributing the population to $|e_1
\rangle$ and $|e_3 \rangle$. Simultaneously, coherence between these states
are created. Such a pulse sequence is simply to prepare dynamics for the later
demonstration of the measurement, but it is not necessarily for real. The
system is initially prepared in state $|g \rangle$.

Since in this work the relaxation channels from highly excited states $|e_2
\rangle$ or $|e_3 \rangle$ to the lowly excited state $|e_1 \rangle$ are
allowed, there are certain probabilities for the wave components of $|e_1
\rangle$ and $|e_2 \rangle$ to jump back to state $|e_1 \rangle$ in each time
step $\mathrm{d} t$, as is considered in MCWF. The rates of the random quantum
jump from states $|e_2 \rangle$ and $|e_3 \rangle$ to $|e_1 \rangle$ during
the evolution of the wave function are $\Gamma_1$ and $\Gamma_2$,
respectively. The quantum jump occurs when $\Gamma_i  | c_i |^2 \mathrm{d} t >
\varepsilon$, where $\varepsilon$ is a uniformly distributed random number in
the range $[0, 1)$. Here we set $\Gamma_1 = \Gamma_2 = 0.007$. Once the jump
occurs, the population of the high energy state is reset to zero, while the
population is transferred to the lower state, and the amplitude of the lower
state is appended by a random phase. In MCWF, the single emission event is
intuitively described by the quantum jump. The single atom may experience
several jump events, leading to the discontinuous "trajectories" of the wave
amplitudes. From the view of a single atom, once the jump occurs at
$t_{\text{j}}$, $c_i$ is no longer continuous. From the view of the ensemble,
the jumping instants $t_{\text{j}}$ vary from atom to atom, leading to
diversified discontinuous trajectories.

For the $s$th single atom, we assume that $c_i^{(s)}$ is the wave function
generated using the MCWF method. Substituting $c_i^{(s)} (t)$ (trajectories)
into Eq. (\ref{eq:Mp-sum}), the amplitude $M_p^{(s)}$ can be calculated. The
partial momentum distribution contributed from sample $s$ is given by $w^{(s)}
(p ; \tau) = | p | | M_p^{(s)} |^2$. The observable of an open system is the
statistical average over all single trajectories. Thus, the final momentum
distribution is the incoherent summation over partial spectra, i.e., $w (p ;
\tau) = \sum_s w^{(s)} (p ; \tau)$. In practice, as shown in this work, 100
trajectories (time-variant $c_i (t)$) are generated and stored in an ensemble,
from which we later calculate the partial momentum distributions $w^{(s)} (p ;
\tau)$, and eventually obtain the final momentum distribution $w (p ; \tau)$.
The true values of the density matrix elements are also evaluated using the
stored $c_i^{(s)}$. The population $\rho_{i  i} = \sum_s | c_i^{(s)} |^2$ and
coherence $\rho_{i  j} = \sum_s (c_i^{(s)})^{\ast} c_j^{(s)}$, as the
benchmark, are compared with the reconstructed values in the main text.

\end{document}